\documentclass[aps,prl,twocolumn,superscriptaddress,showpacs]{revtex4-1}

\usepackage{times,amssymb,amsmath}
\usepackage[pdftex]{graphicx}
\renewcommand{\vec}[1]{\mbox{\boldmath$\mathrm{#1}$}}
\mathchardef\mhyphen="2D
\usepackage{pstricks}
\begin{document}


\title{On the initial stage of quasiparticle decay}
\author{Y. Pavlyukh}
\email[]{yaroslav.pavlyukh@physik.uni-halle.de}
\affiliation{Institut f\"{u}r Physik, Martin-Luther-Universit\"{a}t
  Halle-Wittenberg, 06120 Halle, Germany}
\author{A. Rubio}
\affiliation{Nano-Bio Spectroscopy Group and ETSF Scientific Development Centre,
 Dpto. de F{\'i}sica de Materiales, Universidad del Pa{\'i}s Vasco, 
 CFM CSIC-UPV/EHU-MPC and DIPC, Av. Tolosa 72, E-20018 San Sebasti{\'a}n, Spain}
\affiliation{Fritz-Haber-Institut der Max-Planck-Gesellschaft, Berlin, Germany}
\author{J. Berakdar}
\affiliation{Institut f\"{u}r Physik, Martin-Luther-Universit\"{a}t
  Halle-Wittenberg, 06120 Halle, Germany}\date{June 6, 2011}
\begin{abstract}
The initial stages of the quasiparticle decay in a Fermi liquid are governed by a
time-scale distinct from the scattering rates as derived from the Fermi golden rule
approach.  We show that the initial decay is non-exponential and that it is determined by
the zeroth spectral moment of the electron self-energy. We analyzed numerically a number
of approximations for the self-energy by comparing with exact configuration interaction
calculations for small finite system with fragmented states. A numerically simple approach
for computing the spreading of the quasiparticle states for large systems is devised.
\end{abstract}


\maketitle
 With the recent spectacular advances in light sources and ultrafast spectroscopic methods
it has become possible to trace the quantum dynamics of electronic systems down to the
atto-second time scale (\cite{light-a,light-b,light-c} and references therein).
One of the prime goals is the understanding of the nature of the formation and decay of
electronic states. In this respect, numerical methods are feasible for few-electron
systems only. For condensed matter one has to resort to different concepts such as the
Landau's theory of Fermi liquids~\cite{LandauBook} that describes (low-energy) long-lived
excitations as \emph{quasiparticles} (QP). Due to residual interactions QPs decay in time
generally.  How this decay proceeds in time $t$ is known under certain conditions only: A
QP decays exponentially as $\sim \exp(-\gamma t)$ with the rate constant
$\gamma(\epsilon)\sim \epsilon^2/\epsilon_F$~\cite{PinesNozieres}. Here $\epsilon$ is the
quasiparticle energy and $\epsilon_F$ is the Fermi energy. This, however, holds true only
for times $t\gg 1/\gamma$.  The importance of this restriction is illustrated by the
following: Let us assume an exponential decay at all $t$, thus the QP peak appears in
frequency with a Lorentzian shape.  This means that at the QP energy $\epsilon$ the
spectral function behaves as
$A_{\mathrm{L}}(\omega;\epsilon)=\frac{1}{\pi}\frac{\gamma}{(\omega-\epsilon)^2+\gamma^2}$.
The standard deviation $\sigma^2(\epsilon)$ of the spectral density given by such a
functional form diverges,
\begin{equation}
\sigma^2(\epsilon)=
\int_{-\infty}^{\infty}d\omega
(\omega-\epsilon)^2 A_{\mathrm{L}}(\omega;\epsilon)\rightarrow\infty.
\label{eq:sgm2}
\end{equation}
Explicit calculations for a three-dimensional (3D) homogeneous electron gas (HEG) show
that this divergence is spurious \cite{Vogt2004} and that the zero, first and second
spectral moments are indeed finite. The convergence of the integral Eq.~(\ref{eq:sgm2}) is
governed by the high-frequency behavior of the spectral function.  Thus, the short-time
limit of the single particle Green's function (from which $A_{\mathrm{L}}$ derives) is of
special interest. In this respect quantum-kinetics indicates a quadratic decay in
time~\cite{Bonitz1999}.  Experimentally, atto-second resolution of electronic states in
condensed matter has already called for a careful inspection of this issue~\cite{atto}.

Here we present a spectral function that exhibits the correct short and
long-time behavior, i.e.
\begin{eqnarray}
\frac{d}{dt}A(t;\epsilon)&\stackrel{t\rightarrow0}{\rightarrow}
&-\sigma^2(\epsilon)t\label{eq:shr-tm},\\
A(t;\epsilon)&\stackrel{t\rightarrow\infty}{\rightarrow}&e^{-\gamma t}.\label{eq:lng-tm}
\end{eqnarray}
These equations are exact and can be obtained nonpertubatively from very general physical
considerations~\cite{Haug1996}. An attempt with a similar goal has been undertaken in
Ref.~\cite{Haug1996}, the resulting spectral function, however, violates the sum
rules and has a shape with the spectral moments finite at any order, at variance with
Ref.~\cite{Vogt2004}.  The spectral function given in this work fulfill all sum
rules and comply with the exact short and long time-limits.  The key ingredients are the
imaginary part of the on-shell electron self-energy
$\gamma(\epsilon)=\mathrm{Im}\Sigma(\omega=\epsilon;\epsilon)$ and the decay constant
$\sigma^2(\epsilon)$ as expressed in terms of the zeroth spectral moment of the
self-energy $\Sigma^{(0)}=\frac1\pi\int_{-\infty}^\infty d \omega\,
|\mathrm{Im}\Sigma(\omega;\epsilon)|$.  The approach provides a recipe to compute the
short-time limit of the electron correlation function on the basis of many-particle
perturbation theory. In particular, we demonstrate how the decay constant
(Eq.~\ref{eq:shr-tm}) can be computed diagrammatically to any desired order in the
interaction.  Conceptually, the problem should be addressed by the quantum kinetic
theory. However in this formalism, analytic calculations of the initial stage of the
quasiparticle are not available and numerical approaches rely on further
approximations~\cite{Thygesen2008,Balzer2009,Friesen2009}.

For the decaying part of the  spectral function we  make an ansatz 
\begin{equation}
A(t;\epsilon)=\exp\left(-\gamma(\epsilon) \frac{t^2}{t+\tau(\epsilon)}\right)\label{eq:At}
\end{equation}
which obeys the two limiting cases [Eqs.~(\ref{eq:shr-tm},\ref{eq:lng-tm})] with
$\sigma^2(\epsilon)=2\gamma/\tau$. The analytic form of the Fourier transform of this
function ($A(\omega;\epsilon)=\frac1\pi\int_0^\infty \cos[(\omega-\epsilon) t]
A(t;\epsilon) dt$) is not known, however it is possible to prove that in the frequency
domain it has exactly three finite spectral moments in accordance
with~\cite{Vogt2004}. The large $\omega$ expansion reads
\[
A(\omega;\epsilon)\sim\frac{6}{\pi}\frac{\gamma}{(\omega-\epsilon)^4\tau^2}
\quad\mathrm{as}\quad\omega\rightarrow\infty.
\]
$A(\omega;\epsilon)$ is normalized,~i.e., 0th spectral moment is one~\cite{suppl-mat}.

According to Altshuler \emph{et al.}~\cite{Altshuler1997} the initial stage of the
quasiparticle decay always involves a formation of the two-particle-one-hole state
($2p1h$). The rate of the process is given by the \emph{first collision time} $1/\tau_1$
and is determined by the corresponding Coulomb matrix elements or, in other words, by the
available phase-space (the energy and the momentum must be conserved). The phase-space
also determines in a crucial way the subsequent stages of the QP decay, which results in the
creation of an increasing number of particles and holes, forming either localized or
delocalized states in a Fock space. In the latter scenario the exponential decay is
established after many generations of particles and holes have emerged. From these very
general arguments it is obvious that the exponential decay requires a certain time to
develop, which in our theory is determined by the parameter $\tau(\epsilon)$. This time
parameter certainly exceeds the first collision time ($\tau(\epsilon)\gg\tau_1$) obtained
from the golden rule arguments applied to \emph{the bare Coulomb interaction} (at the
initial stages the screening is not efficient). This indicates that the time
$\tau(\epsilon)$ cannot be obtained from either the bare nor the screened interaction and
is distinct from the relaxation time at the large-time limit ($1/\gamma(\epsilon)$).

\begin{figure*}
\includegraphics[width=0.95\textwidth]{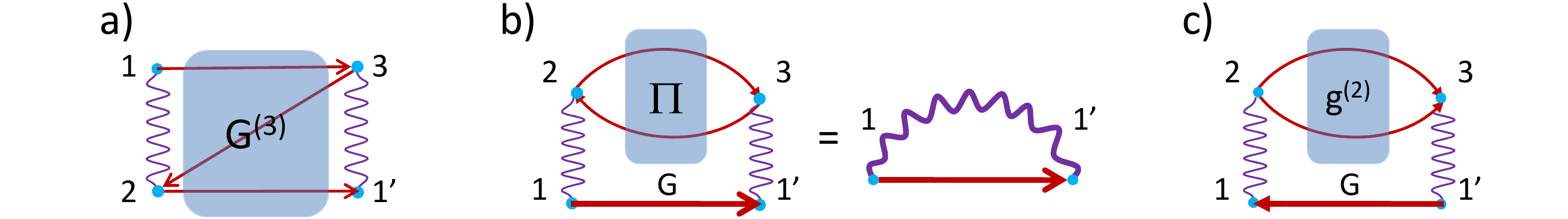}
\caption{Diagrams for the electron self-energy. a) exact expression in terms of the $2p1h$
  response function in which the entrance and exit channels cannot be separated by cutting
  one fermion line; b) $(p\mhyphen h)\mhyphen p$ factorization leading to the $GW$
  approximation; c) $(p\mhyphen p)\mhyphen h$ factorization.\label{fig:Sigma}}
\end{figure*}

To obtain $\tau(\epsilon)$ let us recall the relations between the $n^{th}$ order spectral
moments $\vec{M}^{(n)}$ of the single-particle Green function and that of the self-energy
$\vec{\Sigma}^{(n)}$~\cite{Vogt2004}:
\begin{eqnarray}
\vec{M}^{(0)}=\vec{I}, &\quad&
\vec{\Sigma}_{\infty}=\vec{M}^{(1)}-\vec{\varepsilon},\label{eq:srule1}\\
\vec{\Sigma}^{(0)}&=&\vec{M}^{(2)}-[\vec{M}^{(1)}]^2.\label{eq:srule2}
\end{eqnarray}
 $\vec{\Sigma}_{\infty}$ is the frequency independent real part of the
self-energy~\cite{Schirmer1983}.  These matrix relations directly follow from the Dyson
equation, and can be obtained in any basis (\cite{Vogt2004} used a plane-waves
basis). $\vec \varepsilon$ is a diagonal matrix with the elements given by the
zeroth-order state energies.  For finite systems Hartree-Fock basis states are
appropriate. Writing the matrix of the spectral functions in terms of the imaginary part
of the single-particle Green function
($\vec{A}(\omega)=\frac1\pi|\mathrm{Im}\vec{G}(\omega)|$) and likewise for the spectral
function of the self-energy
($\vec{S}(\omega)=\frac1\pi|\mathrm{Im}\vec{\Sigma}(\omega)|$), and using the
superconvergence theorem~\cite{Altarelli1972} the matrices are cast as frequency
integrals:
\begin{eqnarray}
\vec{M}^{(n)}&=&\int_{-\infty}^{\infty}d\omega\,\omega^n \vec{A}(\omega),\quad
n=0\ldots2,\label{eq:Mn}\\
\vec{\Sigma}^{(0)}&=&\int_{-\infty}^{\infty}d\omega\,\vec{S}(\omega).\label{eq:S1}
\end{eqnarray}
In HF basis $\vec{\Sigma}_{\infty}$ is rather small and is proportional to the difference
of the direct and the exchange Coulomb energy computed with the Hartree-Fock and exact
density matrix, i.~e. related to the deviation of the natural occupations from 1 or
0. Thus, by virtue of Eqs.~(\ref{eq:srule1},\ref{eq:Mn}) we arrive at the conclusion that
the Hartree-Fock energies in the first approximation are given by the center of mass of
the spectral function. Likewise, by using Eqs.~(\ref{eq:srule2},\ref{eq:Mn}) we establish
a formula for the matrix of standard deviations [cf. Eq.~(\ref{eq:sgm2})]:
\begin{equation}
\vec{\sigma}^2=\int_{-\infty}^{\infty}\!\!d\omega\,\vec{S}(\omega).\label{eq:SM2}
\end{equation}
Formally an exact representation of this positively defined matrix can be written in terms
of the six-point response function~\cite{Winter1972} (Fig.~\ref{fig:Sigma}a). Instead, we
will compute the standard deviations for finite systems using its factorizations.

To determine the set-in time of the exponential decay for 3D HEG we apply
Eq.~(\ref{eq:SM2}) in a plane-wave basis to the states close to the Fermi
surface. Calculations of Vogt \emph{et al.}~\cite{Vogt2004} and Farid~\cite{Farid2002}
show that $\sigma^2(\epsilon)$ has two contributions: i) A local, momentum independent and
ii) A non-local, momentum dependent. Their sum remains finite and positive at the Fermi
momentum ($\sigma^2(\epsilon_F)>0$). This tells us that the set-in time for the
exponential decay for quasiparticles in the vicinity of the Fermi surface behaves as
\begin{equation}
 \tau(\epsilon)=\frac{2\gamma(\epsilon)}{\sigma^2(\epsilon)}
\stackrel{\epsilon\rightarrow\epsilon_F}{\sim}\frac{\epsilon^2}{\epsilon_F\sigma^2(\epsilon_F)}.
\end{equation}
The prefactor in front of $\epsilon^2$ can be obtained analytically~\cite{to-be-publ}. We
also note that for HEG the spectral function consist of a quasiparticle peak with the
oscillator strength less the unity (coherent part) surrounded by the satellites
(incoherent part)~\cite{Lundqvist1967}. Our approach goes beyond a description of the
quasiparticle peak: fine detailes of both parts are smeared out preserving, however, the
particle number and having a correct asymptotic behavior. Calculation of higher-order
satellites for comparison requires inclusion of diagrams of higher-orders in the screened
interaction $W$ and is extremely computationally demanding
(\cite{Shirley1996,Takada2001}).  Therefore, we verify the performance of different
approximations for the self-energy by comparing its zeroth moments computed by the
configuration interaction (CI) approach for a \emph{finite electron system}.

As a prototypical system we consider the widely studied Na$_9^+$
cluster~\cite{TDHF,LifeTime,Conserving1}. For the current purpose it is advantageous for
several reasons: i) it contains a small number of electrons making it accessible to full
CI~\cite{SigmaCI,CI-GW}, ii) it can be seen as a generalization of 3D HEG to a finite
number of particles (jellium model ~\cite{Ekardt1984}). The $\vec{\sigma}^2$ matrix can be
computed exactly by exact diagonalization of the many-body Hamiltonian. We use an
algorithm by Olsen \emph{et al.}~\cite{OLSEN1988} based on the graphical unitary group
approach~\cite{KNOWLES1984} for the generation of the restricted active space (RAS) and
full CI Hamiltonians. The calculations are performed for each spin multiplicity separately
using spin-adapted basis functions~\cite{Pauncz}. In terms of the matrix elements of the
creation and annihilation operators the spectral moments are expressed as
\begin{equation}
\vec{M}^{(n)}=\sum_p^{D^{\operatorname{N+1}}}(\varepsilon_p^+)^n\vec{X}^p{\overline{\vec{X}}^p}
+\sum_q^{D^{\operatorname{N-1}}}(\varepsilon_q^-)^n\vec{Y}^q{\overline{\vec{Y}}^q},
\label{eq:Mn-ex}
\end{equation}
where the summation is performed over the Hilbert space of the ionized states (dimension
$D^{\operatorname{N-1}}$) and electron attached states (dimension
$D^{\operatorname{N+1}}$). The matrix elements of electron creation
($\hat{a}_\alpha^\dagger$) and annihilation ($\hat{a}_\alpha$) operators
\begin{equation}
X_\alpha^p=\langle p\mathrm{N+1}|\hat{a}_\alpha^\dagger|0\mathrm{N}\rangle,\quad
Y_\alpha^q=\langle q\mathrm{N-1}|\hat{a}_\alpha|0\mathrm{N}\rangle,\label{eq:XY}
\end{equation}
and the transition energies $\varepsilon_p^+=E_p^{\operatorname{N+1}}-E_0^\mathrm{N}$,
$\varepsilon_q^-=E_0^{\mathrm N}-E_q^{\mathrm N-1}$ are computed from the CI many-body states.

Approximations for the self-energy operator can be obtained from the factorization of the
$2p1h$ six-points function~\cite{Winter1972,Schirmer1983} (Fig.~\ref{fig:Sigma}a). If the
particle-hole ($p\mhyphen h$) Green's function is treated exactly we obtain the so-called
$GW$ approximation~\cite{Onida2002} (Fig.~\ref{fig:Sigma}b). Alternatively, this
approximation can be obtained from the $\Psi[G,W]$ variational energy
functional~\cite{TotEn} expanded in terms of the dressed electron propagator $G$ and the
screened Coulomb interaction $W$. A single diagram of the first order has to be
considered. Finally, one obtains the same functional form by neglecting the three-point
vertex function $\Gamma$ in Hedin's equations~\cite{Hedin}. It should be noted that we do
not perform the self-consistent solution of Hedin's equations; instead, we compute exactly
$G$ and $W$ from the exact one-particle and particle-hole propagators. They are given by
the Lehmann representations in terms of many-body electron
states~\cite{Kobe,Cederbaum1981}. In accordance with the spectral representation of the
self-energy~\cite{SigmaCI} we obtain for the energy-uncertainty:
\begin{equation}
\vec{\Sigma}^{(0),\mathrm{GW}}_{\alpha,\beta}= 4\sum_{n\neq 0}^{D^{\operatorname{N}}}
\sum_p^{D^{\operatorname{N+1}}}\langle\alpha p|n\rangle
\langle n| p\beta\rangle
+(p\mhyphen h)\mhyphen h\mbox{ terms},\label{eq:S0-GW}
\end{equation}
where we introduced a notation for the convolution of the Coulomb matrix elements
$\langle\alpha\beta|\gamma\delta\rangle=\int d(\vec{r}_1\vec{r}_2)\phi_\alpha(\vec{r}_1)
\phi_\beta(\vec{r}_1)\phi_\gamma(\vec{r}_2)\phi_\delta(\vec{r}_2)/|\vec{r}_1-\vec{r}_2|$
with matrix elements of the creation (or annihilation) operators [Eq.~(\ref{eq:XY})] and with
the density matrix elements $Q_{\gamma\delta}^n=\langle
n\mathrm{N}|\hat{a}_\gamma^\dagger\hat{a}_\delta|0\mathrm{N}\rangle$:
\begin{equation}
\langle\alpha p|n\rangle=\sum_{\beta}\sum_{\gamma\delta}X^p_\beta\langle\alpha\beta|\gamma\delta\rangle
Q_{\gamma\delta}^n.\label{eq:me-1}
\end{equation}
Analogically, one obtains an expression for the self-energy using the $(p\mhyphen
p)\mhyphen h$ factorization (Fig.~\ref{fig:Sigma}~c):
\begin{equation}
\vec{\Sigma}^{(0),\mathrm{G^{(2)}}}_{\alpha,\beta}= \sum_{m}^{D^{\operatorname{N+2}}}
\sum_q^{D^{\operatorname{N-1}}}\langle\alpha q|m\rangle\langle m| q\beta\rangle
+(h\mhyphen h)\mhyphen p\mbox{ terms},\label{eq:S0-G2}
\end{equation}
where similar to Eq.~(\ref{eq:me-1}) we define the convolution of
$\langle\alpha\beta|\gamma\delta\rangle$ with the matrix elements of the two creation
(annihilation) operators $P_{\gamma\delta}^m=\langle
n\mathrm{N+2}|\hat{a}_\gamma^\dagger\hat{a}_\delta^\dagger|0\mathrm{N}\rangle$:
\begin{equation}
\langle\alpha q|m\rangle=\sum_{\beta}\sum_{\gamma\delta}Y^q_\beta\langle\alpha\beta|\gamma\delta\rangle
P_{\gamma\delta}^m.\label{eq:me-2}
\end{equation}
Eqs.~(\ref{eq:S0-GW},\ref{eq:S0-G2}) can be thought of as \emph{the Fermi golden rule}
expressions. Since the delta-function ensuring the energy conservation is not present here
the whole expression has a dimension of the energy squared. Starting from
Eqs.~(\ref{eq:S0-GW},\ref{eq:S0-G2}) we further derive a series of simpler
approximations. When the HF Green's function is used in place of $G$ we obtain the
so-called $G^{0}W$ approximation. If, furthermore, the non-interacting excited states are
used to compute $Q_{\gamma\delta}^n$ we obtain the $G^0W^0$ approximation with the
spectral moment:
\begin{equation}
\vec{\Sigma}^{(0),\mathrm{G}^0\mathrm{W}^0}_{\alpha,\beta}=2\sum_{\eta,\gamma,\delta}
\langle\alpha\eta|\gamma\delta\rangle\langle\gamma\delta|\eta\beta\rangle
n_\gamma(1-n_\delta),
\label{eq:S0-G0W0}
\end{equation}
where $n_\gamma$ is the occupation of the single-particle state $\gamma$.

\begin{figure*}
\makebox[0.33\textwidth][l]{\Large a)}
\makebox[0.32\textwidth][l]{\Large b)}
\makebox[0.3\textwidth][l]{\Large c)}
\includegraphics[width=0.33\textwidth]{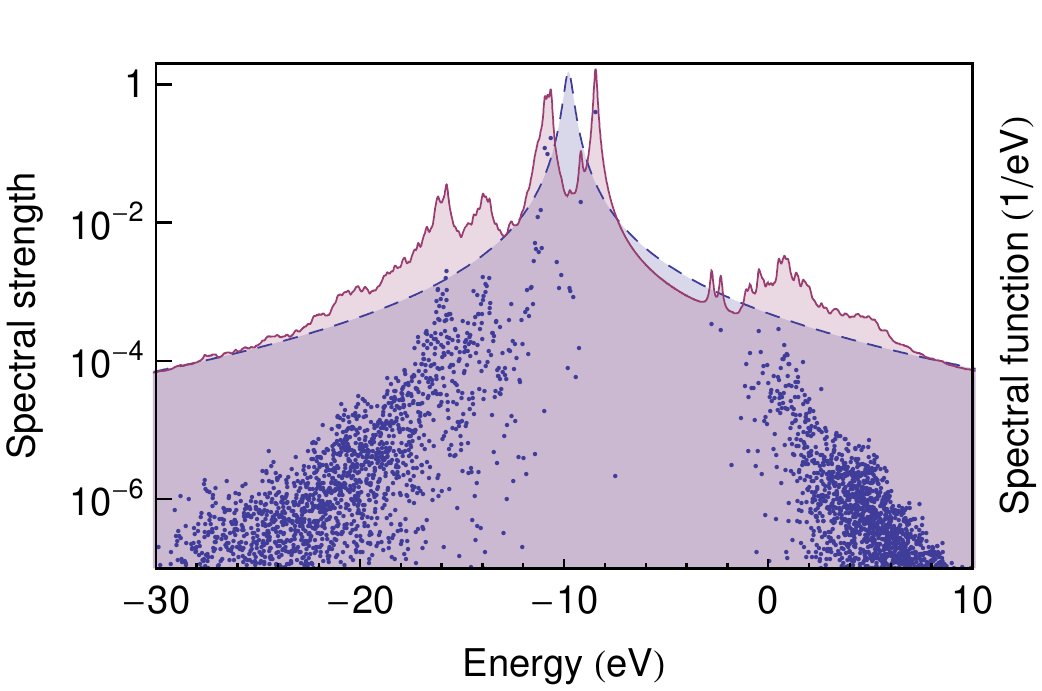}
\includegraphics[width=0.32\textwidth]{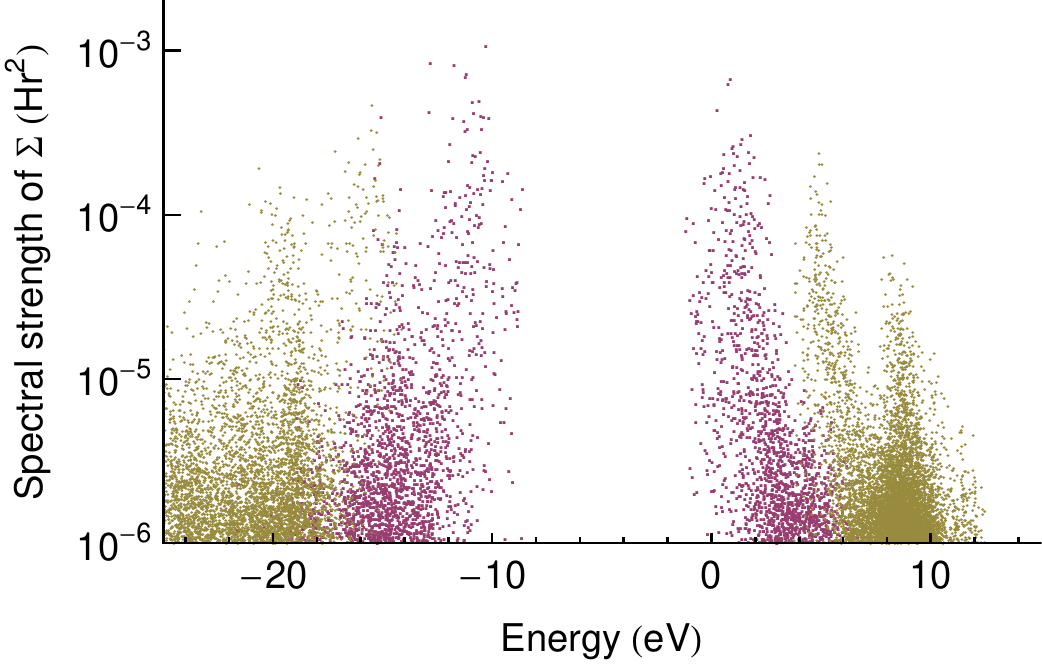}
\includegraphics[width=0.3\textwidth]{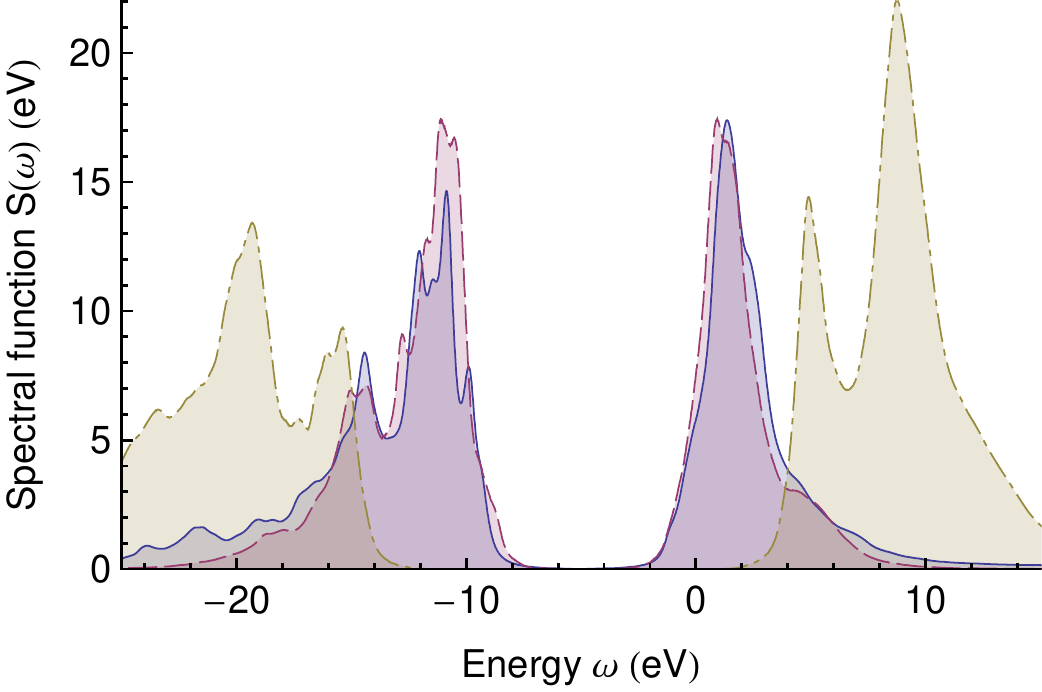}
\caption{(Color online) CI calculations for the Na$_9^+$ cluster: a) The exact (solid) and
  model (dashed) spectral functions of lowest valence state. The dots denote the weights
  (squares of the matrix elements, Eq.~(\ref{eq:XY})) of the exact spectral function; b)
  The spectral strength of the self-energy in $(p\mhyphen h)\mhyphen p$ (red) and
  $(p\mhyphen p)\mhyphen h$ (yellow) factorizations; c) The spectral function
  ($\mathrm{Tr}[\vec S(\omega)]$) of the exact (solid), $(p\mhyphen h)\mhyphen p$ (dashed) and
  $(p\mhyphen p)\mhyphen h$ (dash-dotted) self-energies.
\label{fig:sline}}
\end{figure*}

\begin{figure}[h!]
\includegraphics[width=\columnwidth]{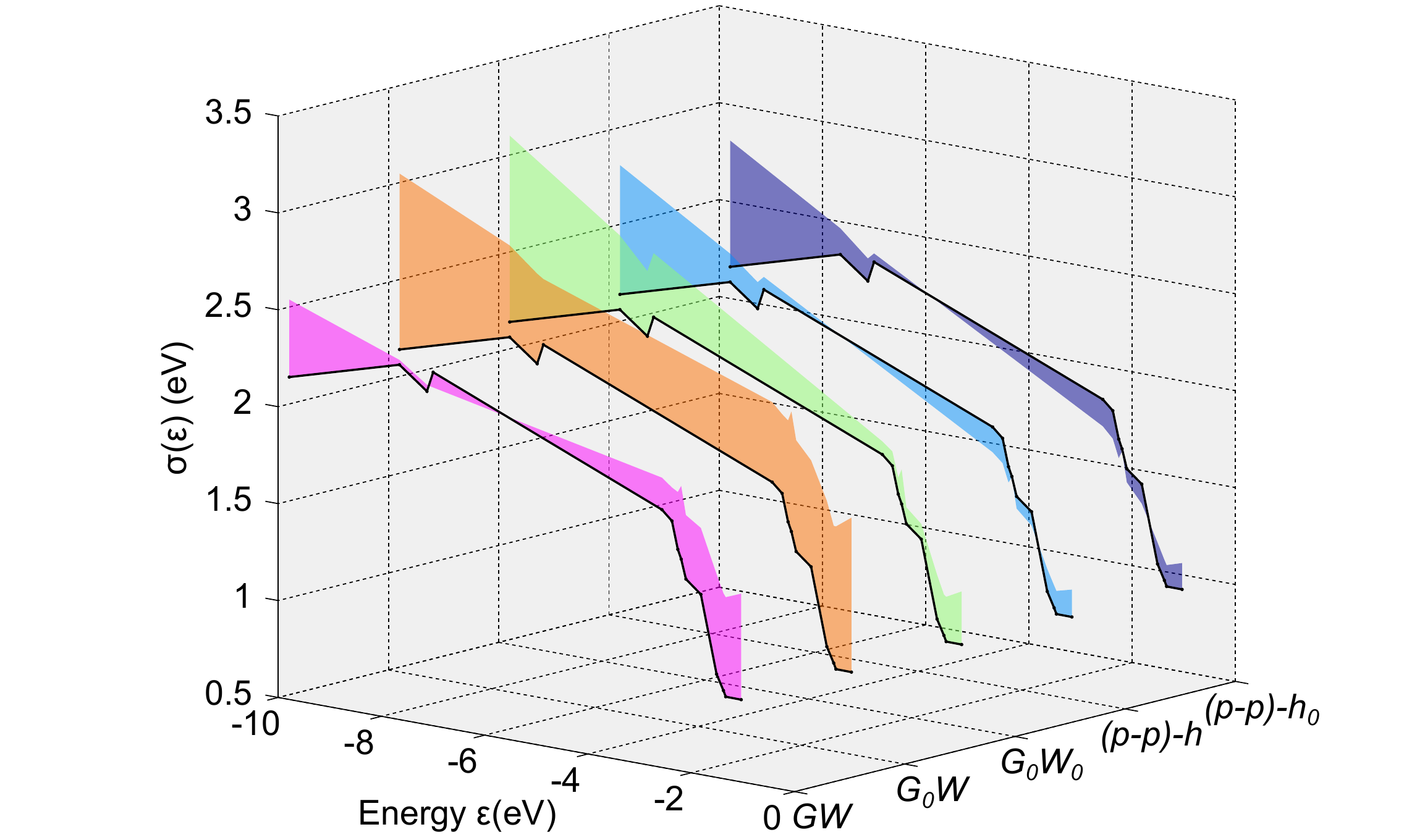}
\caption{(Color online) Energy-uncertainty for the Na$_9^+$ states. Shaded area denotes
  deviation of the approximate theories from the exact. 
  \label{fig:sgm}}
\end{figure}

For the Na$_9^+$ cluster we computed the spectral moments
(Eqs.~\ref{eq:Mn-ex},\ref{eq:S0-GW},\ref{eq:S0-G2}) by exact diagonalization of the
many-body Hamiltonian. The exact spectral function (solid line, Fig.~\ref{fig:sline}a) of
the lowest valence state ($\varepsilon_1^{\operatorname{HF}}=9.786$~eV) is fragmented (two
major peaks) and has multiple satellites. Despite this fact the model spectral function
centered at the HF energy approaches the exact one in a large range of energies.  The
exact energy-uncertainty from the first two spectral moments of the spectral function
(Eq.~\ref{eq:Mn-ex}) is compared with the expressions resulting from the approximation for
the self-energy (Eqs.~\ref{eq:S0-G0W0},\ref{eq:S0-GW},\ref{eq:S0-G2})
(Fig.~\ref{fig:sgm}). Corresponding self-energy spectral functions and weights are in
Fig.~\ref{fig:sline}(b,c). Generally, $GW$ self energy yields results superior to other
approximations. Despite the fact that $(p\mhyphen p)\mhyphen h$ factorization performs bad
for $\vec S(\omega)$ (large energy gap in the case of finite systems) the spectral moments
are close to those of $G^0W^0$ approximation.

Summarizing, we presented a form of the quasiparticle line-shape that reflects the correct
short and long time-limits of the single-particle Green function and, thus, can be used to
parametrize evolving in time electronic structure (e.~g., attosecond time-resolved
photoemission). The spectral function also explicitely enters a description of a number of
static processes (e.~g. core state x-ray photoemission). Thus, the experimentally observed
form of the Fermi edge singularity will be affected by both the finite life-time of the
core state $1/\gamma$ as well as by the corresponding set-in time $\tau$~\cite{suppl-mat}.

In the case of 3D HEG the spectral function describes both the coherent and incoherent
parts. We show that the set-in time $\tau(\epsilon)$ vanishes as $\epsilon^2$ in the
vicinity of the Fermi surface. In the case of finite systems the CI method enables us to
compute the energy uncertainties using a number of diagrammatic approximations. Our
simulations indicate that accurate results are obtained even by neglecting the
three-particle vertex, however, further verifications are needed for extended systems.

The work is supported by DFG-SFB762 [YP,~JB], and by the Spanish MEC
(FIS2010-21282-C02-01), ACI-promciona project (ACI2009-1036), ‘‘Grupos Consolidados
UPV/EHU del Gobierno Vasco’’ (IT-319-07), the EU e-I3 ETSF project no. 211956 [AR].

\begin{thebibliography}{36}
\bibitem [{to-()}]{to-be-publ}%
  \BibitemOpen
  \href@noop {} {}\bibinfo {note} {Will be published elsewhere.}\BibitemShut
  {Stop}%
\end{thebibliography}
\end{document}



\title{On the initial stage of quasiparticle decay\\-- Supplementary Material --}
\author{Y. Pavlyukh}
\affiliation{Institut f\"{u}r Physik, Martin-Luther-Universit\"{a}t
  Halle-Wittenberg, 06120 Halle, Germany}
\author{A. Rubio}
\affiliation{Nano-Bio Spectroscopy Group and ETSF Scientific Development Centre,
 Dpto. de F{\'i}sica de Materiales, Universidad del Pa{\'i}s Vasco, 
 CFM CSIC-UPV/EHU-MPC and DIPC, Av. Tolosa 72, E-20018 San Sebasti{\'a}n, Spain}
\affiliation{Fritz-Haber-Institut der Max-Planck-Gesellschaft, Berlin, Germany}
\author{J. Berakdar}
\affiliation{Institut f\"{u}r Physik, Martin-Luther-Universit\"{a}t
  Halle-Wittenberg, 06120 Halle, Germany}\date{\today}
\begin{abstract}
\end{abstract}
\maketitle
\subsection{General properties}
The spectral function of a fermionic many-body quantum system is defined in its most
general form as the overlap of a particle (hole) states:
\[
2\pi S(t-t';\epsilon)=\langle \psi(t;\epsilon)|\psi(t';\epsilon)\rangle.
\]
The states emerge by creating a particle for $\epsilon>\epsilon_F$ or a hole with
$\epsilon\le\epsilon_F$ and are not the eigenstates of the system in general and, thus,
decay in time. It is convenient to represent the spectral function as a product of a
noninteracting oscillatory part and a decaying part $A(t;\epsilon)$:
\[
S(t;\epsilon)=\frac{e^{-i\epsilon t}}{2\pi} A(t;\epsilon).
\]
 For the latter we make an ansatz
\begin{equation}
A(t;\epsilon)=\exp\left(-\gamma(\epsilon) \frac{t^2}{t+\tau(\epsilon)}\right).\label{eq:At}
\end{equation}

\begin{figure}[b!]
\includegraphics[width=0.95\columnwidth]{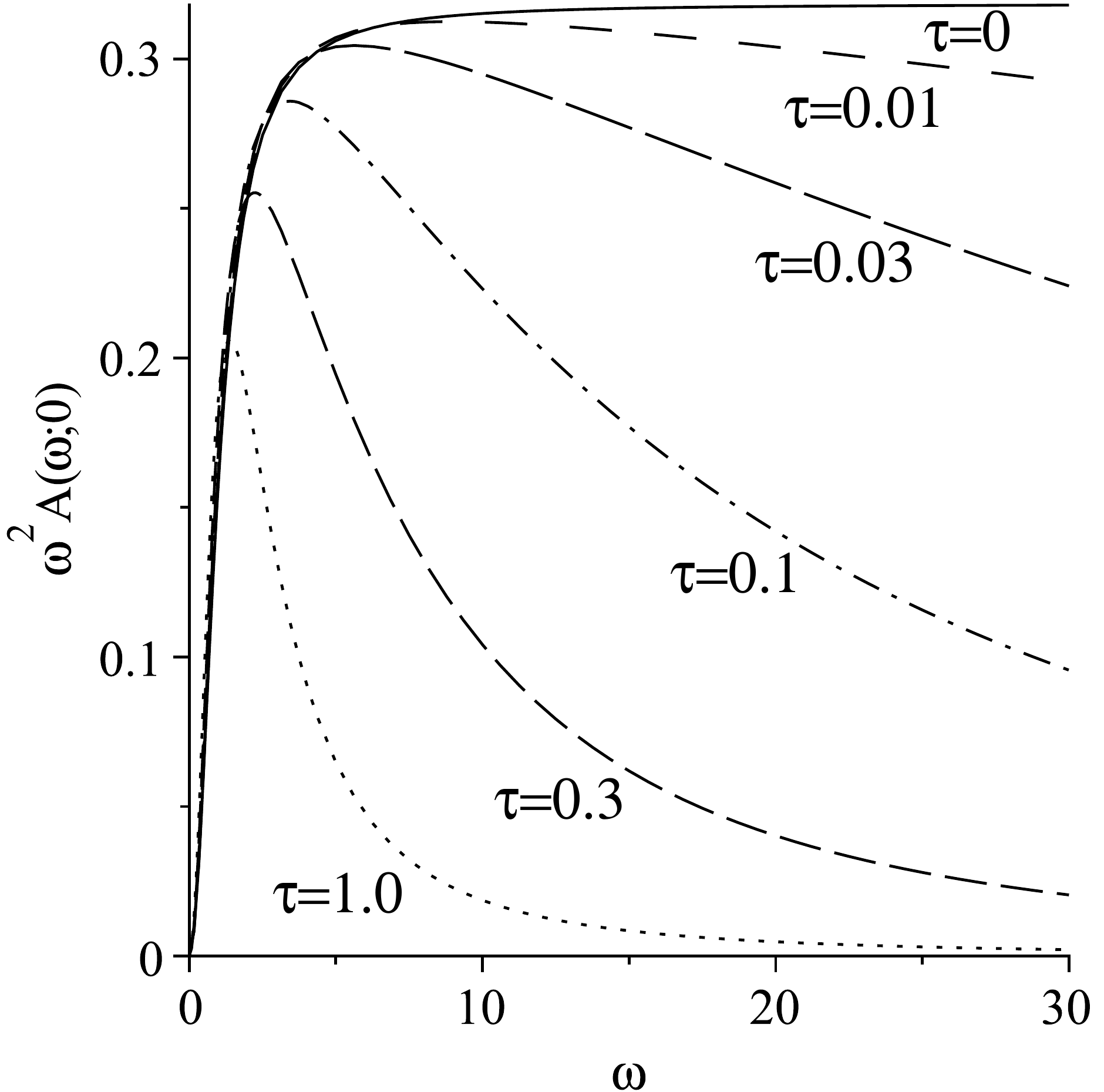}
\caption{$\omega^2A(\omega;0)$ for the model spectral function defined by
  Eq.~(\ref{eq:At}) for the rate constant $\gamma=1.0$ and different values of the set-in
  time. \label{fig:Aw}}
\end{figure}

The Fourier transform of the spectral function can be written as follows:
\begin{eqnarray}
A(\omega;\epsilon)&=&\int_{-\infty}^{\infty}dt\,e^{i\omega t}S(t;\epsilon)\nonumber\\
&=&\frac1\pi\int_0^\infty \cos[(\omega-\epsilon) t] A(t;\epsilon) dt.\label{eq:FT}
\end{eqnarray}
For $A(t;\epsilon)$ given by Eq.~(\ref{eq:At}) the analytic form of the Fourier transform
is not known, but can easily be obtained numerically [Fig.~(\ref{fig:Aw})]. Its odd
spectral moments are zero because of symmetry consideration
$A(t;\epsilon)=A(-t;\epsilon)$:
\[
\bar{M}^{(2k-1)}(\epsilon)=\int_{-\infty}^{\infty}d\omega\,(\omega-\epsilon)^{2k-1}
A(\omega;\epsilon)=0.
\]
Thus $\bar{M}^{(1)}(\epsilon)=M^{(1)}(\epsilon)-\epsilon=0$.  The even spectral moments can
analytically be obtained from the derivatives of $A(t;\epsilon)$ at $t=0$:
\[
\bar{M}^{(2k)}(\epsilon)=\int_{-\infty}^{\infty}d\omega\,(\omega-\epsilon)^{2k}
A(\omega;\epsilon)= (-1)^k\lim_{t\rightarrow0}A^{(2k)}(t;\epsilon).
\]
This leads to $\bar{M}^{(0)}(\epsilon)=M^{(0)}(\epsilon)=1$ (normalization condition), and
$\bar{M}^{(2)}(\epsilon)=
M^{(2)}(\epsilon)-[M^{1}(\epsilon)]^2=\frac{2\gamma(\epsilon)}{\tau(\epsilon)}$. Higher
spectral moments diverge because of the discontinuity of the derivative of $A(t;\epsilon)$
at $t=0$. This competes the prove that in the frequency domain it has exactly three finite
spectral moments in accordance with~\cite{Vogt2004}.

The asymptotic large $\omega$ expansion can readily be obtained by integrating
Eq.~(\ref{eq:FT}) by parts :
\[
A(\omega;\epsilon)\sim\frac{6}{\pi}\frac{\gamma}{(\omega-\epsilon)^4\tau^2}
\quad\mathrm{as}\quad\omega\rightarrow\infty.
\]
\begin{figure*}[t!]
\includegraphics[width=0.95\columnwidth]{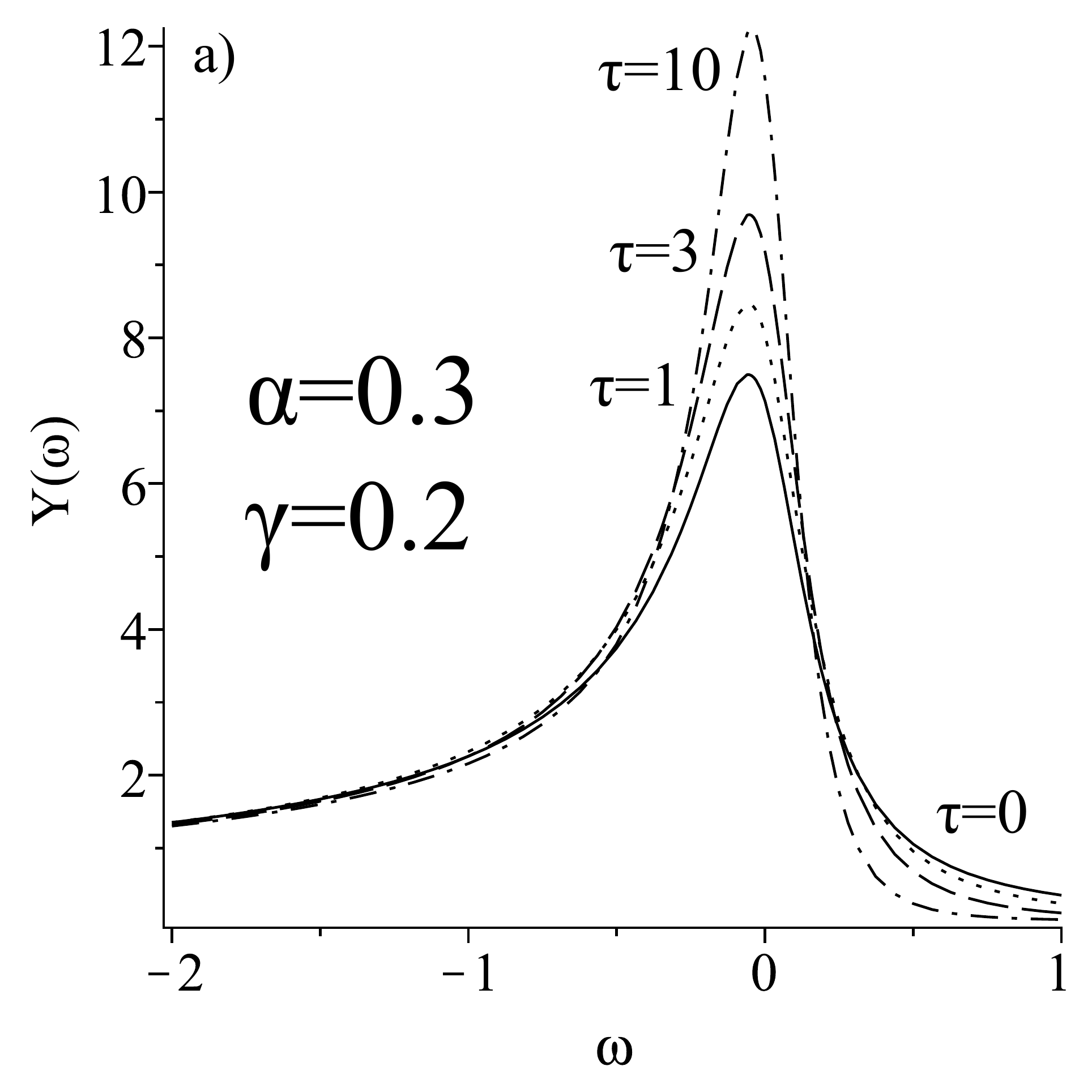}\hfill
\includegraphics[width=0.95\columnwidth]{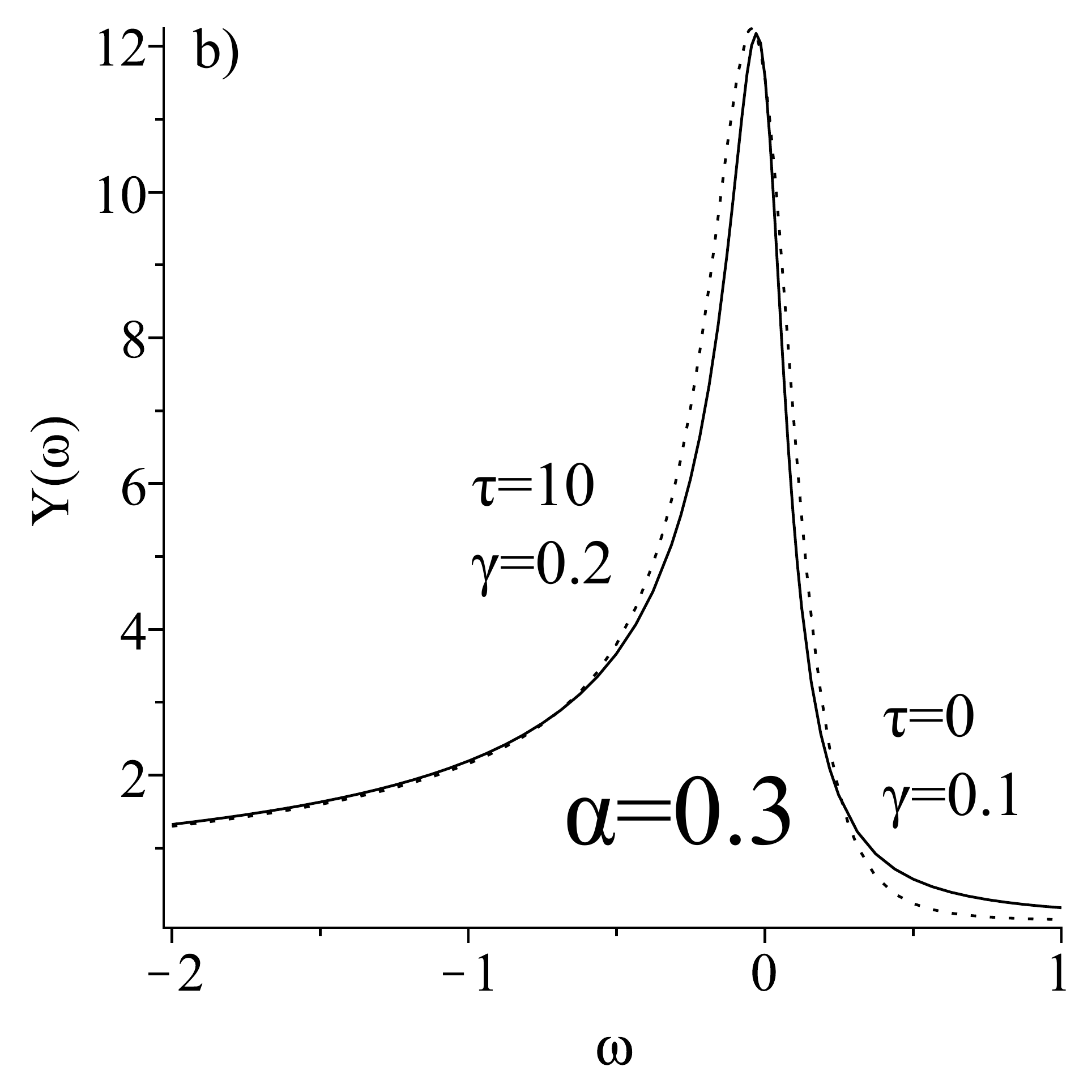}
\caption{a) Line-shapes for the x-ray photoemission from core states with finite core
  state broadenings $\gamma$ and different set-in times $\tau$. $\tau=0$ (solid line)
  corresponds to the Doniach and \u{S}unji\'{c} (DS) line-shape. b) The effect of finite
  set-in time (dotted) can results in a XPS profile experimentally indistinguishable from
  the DS line-shape with a substantially reduced broadening.\label{fig:Yw}}
\end{figure*}
\subsection{Fermi edge singularities}
The proposed spectral function can be applied to obtain more accurate descriptions of the
x-ray photoemission (XPS) line shapes from core states. The theory describing the
many-electron response in XPS originated in works of
Anderson~\cite{Anderson1967a,Anderson1967b}, Mahan~\cite{Mahan1967}, Nozi{\`e}res and
DeDominicis~\cite{Nozieres1969}. It was shown that a creation of an infinite number of
electron-hole pairs accompanies the photoemission event. The resulting electron spectrum
for an infinitely-long-lived core hole is given by
$\frac{2\pi}{\Gamma(\alpha)}\frac{\theta(-\omega)}{\omega^{1-\alpha}}$, where
$0<\alpha<\frac12$ is called the Anderson singularity index and can be given in terms of
the scattering phase shifts $\delta_l$ as
\[
\alpha=2\sum_{l}(2l+1)\left(\frac{\delta_l}{\pi}\right)^2.
\]
For brevity we set the origin of the energy scale at the no-loss position of the
hole-state.

This approach is oversimplified and a proper treatment requires a finite lifetime of the
core holes to be taken into account. Convolution of the singular line-shape with a simple
decay function $\exp(-\gamma t)$ yields the well-known Doniach and
\u{S}unji\'{c}\cite{Doniach1970} profile:
\[
Y^0(\omega)=\frac{2\Gamma(1-\alpha)}{(\omega^2+\gamma^2)^{\frac{1-\alpha}2}}
\cos\left[\frac12\pi\alpha+(1-\alpha)\arctan\left(\frac{\omega}{\gamma}\right)\right],
\]
where $\Gamma$ is the gamma function.

It is clear that even this modification is insufficient as it is based on the wrong
assumption of the hole's spectral function. Therefore, we propose a modified line shape
based on the presented spectral function with correct asymptotic behavior.
\begin{equation}
Y(\omega)=\lim_{\delta\rightarrow+0}
\int_{-\infty}^\infty \exp\left(-\frac{\gamma t^2}{\tau+|t|}\right)e^{i\omega t}\frac{dt}{(\delta-it)^\alpha},
\end{equation}
where infinitesimally small positive $\delta$ was introduced to select a branch of the
multivalued $1/(it)^\alpha$ function and to facilitate the numerical integration
(Fig.~\ref{fig:Yw}).

A comparison of XPS line-shapes for different set-in times (Fig.~\ref{fig:Yw},~b) reveals
a similarity between the effect of finite set-in time and that of the reduced
broadening. In other words, while $\gamma$ leads to the broadening of spectral lines the
presence of non-zero set-in time $\tau$ has an opposite effect as evidenced by
$\bar{M}^{(2)}(\epsilon)= \frac{2\gamma(\epsilon)}{\tau(\epsilon)}$. This observation
might have a strong impact on the interpretation of the experimental XPS spectra.

We also note a different approach leading to a modified spectral line-shape, the
assumption of the frequency dependent scattering phase as was recently implemented for
low-dimensional systems by Mkhitaryan and Raikh~\cite{Mkhitaryan2011}.